# Two-photon quantum walk in a multimode fiber

Hugo Defienne[1], Marco Barbieri[2], Ian A. Walmsley[3], Brian J. Smith[3], Sylvain Gigan[1]

[1]Laboratoire Kastler Brossel, ENS-PSL Research University, CNRS, UPMC-Sorbonne universités, Collège de France ; 24 rue Lhomond, F-75005 Paris, France
[2]Università degli Studi Roma Tre, Via della Vasca Navale 84, 00146, Rome, Italy
[3]Clarendon Laboratory, Department of Physics, University of Oxford OX1 3PU, United Kingdom

SUMMARY SENTENCE

Control of a two-photon quantum walk in a complex multimode system by wavefront shaping.

ABSTRACT

Multi-photon propagation in connected structures – a quantum walk – offers the potential for simulating complex physical systems and provides a route to universal quantum computation. Increasing the complexity of quantum photonic networks where the walk occurs is essential for many applications. Here, we implement a quantum walk of indistinguishable photon pairs in a multimode fiber supporting 380 modes. Using wavefront shaping, we control the propagation of the two-photon state through the fiber in which all modes are coupled. Excitation of arbitrary output modes of the system is realized by controlling classical and quantum interferences. This experiment demonstrates a highly multimode platform for multi-photon interference experiments and provides a powerful method to program a general high-dimensional multiport optical circuit. This work paves the way for the next generation of photonic devices for quantum simulation, computing and communication.

INTRODUCTION

In a random walk process the walker chooses which path to take based upon the toss of a coin. Quantum walkers also randomly choose which path to follow, but maintain coherence over the paths taken. This simple phenomenon lies at the core of simulating condensed matter systems, quantum-enhanced search algorithms, and universal models of quantum computation (1,2). Single walker quantum dynamics, which can be understood in terms of classical wave evolution, have been explored by utilizing several physical platforms including atomic systems (3,4) and photonics (5-10). Quantum walks and their applications become more complex with additional walkers and cannot be understood using only classical wave propagation. For example quantum walks with entangled photons (11) have led to simulation of Anderson localization and Fano resonances mimicking fermionic-like systems (12). Further attention has been devoted to the computational complexity of multi-photon quantum walks, which has the potential to demonstrate an undisputed quantum advantage through the boson-sampling algorithm (13–19). Integrated optical circuits permitting nearest neighbor coupling between a few tens of modes have been the key technology utilized in the nascent field of multi-photon quantum walks (20-22). Here we explore an alternative route to high-dimensional mode coupling by harnessing the multimode nature of random media. Coherent wave propagation through a disordered medium results in a speckle pattern arising from the highly-complex multiple-scattering process. Propagation in such complex environment represent a quantum walk, since each scattering event redistributes the photons in different optical modes. While the propagation of non-classical states of light through such media has been explored theoretically and experimentally (23–27), these investigations have yet to realize to full potential of quantum walks, which requires the ability to prepare arbitrary input states of the walkers. A paradigm shift has occurred in the last few years in which digital wavefront shaping has emerged as a powerful approach to manipulate light propagation through complex media (28). This has made significant impacts across a diverse range of research including biomedical imaging, optical sensing and telecommunications (29–31).

While controlled propagation of single-photon states through random media has been recently demonstrated (32–34), scaling this approach to multi-photon states is challenging due to the exceptionally high number of spatial and spectral modes of a diffusive medium. Multimode fiber (MMF) provides a platform that delivers a sufficiently large number of coupled modes (41) to demonstrate a new regime of modal capacity for quantum light, yet the number of modes is small enough to enable controllable propagation using techniques developed with classical light. Strong and complex multimode coupling in the fiber arises from interference between a well-defined number of spatial modes supported by the fiber. Moreover, the use of MMF for quantum applications is enhanced by the near-lossless optical propagation along the fiber. In this system, photons are continuously redirected during their propagation through the coupling between modes, thus undergoing a continuous-time quantum walk (11). Here we marry the highly multimode coupling power of MMF with wavefront shaping techniques to realize a programmable linear-optical network for quantum optical technologies. As an initial demonstration of this approach, we show that two-photon propagation through this highly multimode device can be controlled by shaping the wavefront of each photon on the input of the MMF.

**RESULTS**

Here, we report experimental results of two-photon quantum walks experiment using a 11-cm graded index multimode fiber with a 50 µm core diameter supporting approximately 380 transverse spatial and polarization modes at 810 nm. For our investigation we employ the experimental setup depicted in Fig. 1. Two orthogonally-polarized degenerate narrowband photons derived from a spontaneous parametric down conversion (SPDC) source (see methods) are launched into a MMF. Prior to entering the MMF each photon is individually shaped by spatial light modulators (SLM V and SLM H) to control its transverse spatial profile. Varying the path length difference δ between the individual photon paths by means of a translation stage adjusts their relative arrival time to the MMF. This enables partial control over the distinguishability of the two photons (see Supplementary Materials). While light propagates through the fiber, polarization and spatial modes mix and an unpolarized speckle pattern is observed at the output. The emerging light from the MMF is polarized by means of a polarizing beam splitter (PBS), and then analyzed by either an imaging electron-multiplied charged-coupled device (EMCCD), or a pair of avalanche photodiodes (APD) coupled to polarization-maintaining single-mode fibers (SMFs), mounted on translation stages.

**Reconstruction of a two-photon transmission matrix**

To demonstrate the capacity of a MMF for multimode, multi-photon quantum optics, we first characterize the propagation of photon pairs by recording the two-photon transmission matrix (TTM) of the system.

Different two-photon input states are prepared by programming the SLMs to excite different transverse input modes of the MMF (Fig. 2a). Direct measurements made using the EMCCD camera presented on Fig. 2b show how each photon is delocalized over approximately 50 independent output speckle grains. The overall intensity distribution is the incoherent sum of each individual photon intensity profile and is independent of the indistinguishability of the photons. Intensity-correlations are measured by coincidence detection events between the two output SMFs, which can positioned at two different locations in the output image plane denoted $\{|X_1\rangle, |X_2\rangle\}$ and $\{|Y_1\rangle, |Y_2\rangle\}$ for fiber F1 and fiber F2 respectively. We measure a $16 \times 4$ coincidence matrix in Fig. 2.b by scanning 4 orthogonal input transverse spatial states for each photon, set by the SLMs, and measure the corresponding coincidence rates at four pairs of output fiber positions. The so-called TTM matrix reconstructed here characterizes propagation of photons-pairs through the MMF between the selected states taking into account classical and quantum interferences effects occurring in the fiber.

The measured intensity correlations are modified when the distinguishability of the two photons is adjusted from indistinguishable ($\delta = 0$) to distinguishable ($\delta=0.4$mm), while each individual intensity profile remains unchanged. This signature of two-photon quantum interference is analogous to the Hong-Ou-Mandel effect (35). Quantitative analysis of this two-photon inference is conducted by calculating the non-classical contrast for each input-output configuration (Fig.2d.). The non-classical contrast is defined as $C = (R_{\delta=0} - R_{\delta=0.4mm})/R_{\delta=0.4mm}$, where $R_{\delta=0}$ ($R_{\delta=0.4mm}$) is two-photon coincidence rate at zero (0.4 mm) path length difference between input photons ; the classical bound on the magnitude contrast is 0.5. These results demonstrate that the quantum coherence between the photons is robust under propagation through the MMF, which acts as a high-dimensional multimode platform for quantum interference.

**Control of photon-pairs propagation**
To control the multi-photon interference in the MMF we adopt the wavefront shaping technique employed for imaging through opaque systems that utilizes phase-only SLMs (28-31, 36). This technique exploits the strong and complex mode mixing occurring inside the disordered system to manipulate the classical field at the output. The approach to control the propagation of classical light through a multimode scattering medium in (36) can be generalized to control the propagation of two photons by use of the two-photon transmission matrix (TTM) associated with the random medium. The coincidence matrix reported in Fig. 2 represents direct estimates of a small subset of 64 elements of the TTM. However, given the large number of modes supported by the MMF, approximately 380, a more convenient approach is to calculate the TTM from the transmission matrix (TM) of the MMF, which can be measured with greater precision using classical light (see Methods). Following this approach using the calculated TTM, two SLMs configurations are found to optimize the coincidence rate of the targeted state $\hat{a}^+_{|X_2\rangle}\hat{a}^+_{|Y_3\rangle}|0\rangle$, where $|X_2\rangle$ and $|Y_3\rangle$ are two arbitrary chosen positions in the output plane.

In the first configuration the SLMs are programmed to focus photon H at output position $|X_2\rangle$, and, independently, photon V at output position $|Y_3\rangle$. The EMCCD camera images (Fig.3.a1) show strong localization of the photons at the two targeted positions. The focusing process is confirmed with the intensity correlations measured by the APDs coupled to SMFs, where the coincidence pattern shows a pronounced spike in the coincidence count rate when the collection fibers are set at these positions (Fig.3.a2 and a3). Since each photon is directed independently to a different output, we do not observe significant changes in the coincidence rate when moving from the distinguishable ($\delta = 0.4$mm) to the indistinguishable case ($\delta = 0$).

In the second case each photon is prepared in a superposition of these two output positions (Fig.3b1). SLM H (SLM V) is programmed to direct photon H (V) in a superposition of output states $|X_2\rangle$ and $|Y_3\rangle$ with a controlled relative phase $\varphi_H$ ($\varphi_V$), i.e $|X_2\rangle + e^{i\varphi_H}|Y_3\rangle$ ($|X_2\rangle + e^{i\varphi_V}|Y_3\rangle$). The EMCCD camera images (Fig.3b2) confirm that each photon is delocalized on the two targeted output spatial regions. In this configuration there are multiple paths that can lead to coincidence detection of a photon at output positions $|X_2\rangle$ and $|Y_3\rangle$, in which photon H arrives at $|X_2\rangle$ and V at $|Y_3\rangle$ or vice versa. If these outcomes are indistinguishable, then quantum interference between the multiple paths occurs. When the SLM sets the phase condition $\varphi_H = \varphi_V$, the multiple paths will interfere constructively. As shown on Fig 3b3, we observe an increase of the coincidence rate by 72% for indistinguishable photons ($\delta = 0$) compared to distinguishable photons ($\delta = 0.4$mm), while single counts remain unaffected. Our method enables the output two-photon state to be prepared in a well-controlled superposition of the two targeted output spatial states by exploiting both classical and quantum interferences.

Complete control over quantum interference between the two photons is presented in Fig 4. Non-classical contrast measurements performed for 64 different relative phases settings ($\varphi_H, \varphi_V$) are presented in Fig.4a. Photon pairs interfere constructively, leading to a photon bunching effect, when $\varphi_H = \varphi_V$, and destructively when $\varphi_H = \varphi_V \pm \pi$. By scanning the path length difference between the two photons for three specific phase settings, $(\varphi_H, \varphi_V) \in \{(0,0), (0,\pi), (0,\pi/2)\}$, we retrieve three HOM-like plots (Fig. 4b) displaying a peak (green), a dip (red), and flat line (blue). We thus demonstrate the possibility to distribute both photons into two arbitrary output states of the system with a complete control over the coupling parameters.

## DISCUSSION

Platforms for multimode interference of multi-photon states hold promise for a variety of quantum applications. A key experimental challenge with this approach resides in scaling up the number of modes and photons involved in such experiments. Here we present a technique that uses knowledge of the transmission matrix of a multimode fiber combined with wavefront shaping methods to demonstrate control of two-photon interference in set of selected output spatial states. Our results show that this system has the potential for a programmable multimode optical network (37), establishing multimode fibers as a reconfigurable, high-dimensional platform for multi-photon quantum walks. Low losses, stability and scalability of this device hold promise for realization of quantum walks in regimes where classical verification becomes challenging. Moreover, photonic lanterns (38) could permit simple and efficient interfacing of the MMF with a large array of detectors. The use of broadband non-classical sources combined with recently developed spatio-temporal wavefront shaping techniques (39,40) provides additional degrees of freedom to increase the number of modes that can be addressed and extend the possibilities for quantum information processing.

## METHODS

*Experimental details*
Photon-pairs are generated by a type-II SPDC process. A 10 mm periodically poled potassium titanyl phosphate (PPKTP) crystal pumped with a 25 mW continuous-wave laser diode at 405 nm produces pairs of frequency-degenerate photons at 810 nm. Both photons are spectrally filtered using narrowband interference filters centered at 810 nm (1 nm full width at half maximum bandwidth). A two-photon interference contrast of 86% is measured by performing a Hong-Ou-Mandel interference experiment using a balanced beam splitter (see Supplementary Materials).
The classical source used to record the TM is a superluminescent diode (SLED) with a 20nm bandwidth spectrum centered at 810 nm. The source is filtered with the same narrowband filter (FWHM = 1nm) used with the SPDC source. Both sources are coupled to the wavefront shaping apparatus using PMSFMs and can be easily swapped.
The SLM a phase only liquid-crystal SLM subdivided in two independent parts denoted SLM H and SLM V. Each part has an active area of $8 \times 9$ mm$^2$ and a resolution of $960 \times 1080$ pixels.
The experiments use a graded-index MMF with 50µm core diameter and 11 cm length (Reference number: GIF50C, from Thorlabs). This MMF carries about 380 optical modes. Its properties have been chosen to neglect temporal dispersion effects and light propagation can be considered quasi-monochromatic (see Suppplentary Materials).

*TM measurement*
The input face of the MMF and the SLM plane are Fourier conjugated by a lens of focal length $f = 20$ mm (Fig.1). When a phase ramp is programmed on the SLM, a diffraction-limited spot of light is focused at a specific position on the MMF input face. At the output of the MMF, intensity speckle patterns are acquired using both an EMCCD camera and two PMSMFs coupled to APDs. The TM is reconstructed by recording output fields with a phase-stepping holographic method for different positions of the focused spot. Each input mode is then corresponds to a certain position of the focused spot - or in an equivalent way a certain phase ramp pattern on the SLM - and an output mode is defined either as the spatial mode of one of the PMSMFs or as an EMCCD camera pixel. In this experiment, the TM is measured for both input polarizations by programming 180 spatial modes on SLM H and 190 spatial modes on SLM V, leading to a total of $N$=370 modes. At the end of the process, the TM is projected onto the SLM pixel basis using a basis change matrix multiplication.

*Controlling photon-pairs propagation using the TTM*

The TTM matrix is calculated from the measured TM. By analogy to the method described in (36), the transpose conjugate of the TTM is used as an inverse operator to determine the two-photon input field that allows focusing photon-pairs into output state $\hat{a}^+_{|X_2\rangle}\hat{a}^+_{|Y_3\rangle}|0\rangle$. The two SLM configurations presented in Fig.3 are equivalent solutions of the inverse process maximizing the coincidence rate in $\hat{a}^+_{|X_2\rangle}\hat{a}^+_{|Y_3\rangle}|0\rangle$ under our experimental constraints (see Supplementary Materials).

SUPPLEMENTARY MATERIALS

Fig. S1. Characterization of the photon-pairs indistinguishability by Hong-Ou-Mandel experiment
Fig. S2. Dispersion characterization through the MMF
Fig. S3. Statistical analysis of experimental data represented on Fig.2
Fig. S4. Statistical analysis of experimental data presented on Fig.4

AKNOWLEDGEMENTS

All data needed to evaluate the conclusions in the paper are present in the paper and/or the Supplementary Materials. Additional data related to this paper may be requested from the authors. The authors thank Etienne Werly, Beatrice Chatel and Benoit Chalopin for fruitful discussions, David Martina and Justin Spring for technical support. This work was funded by the European Research Council (grant no. 278025) and partially supported by the Rita Levi-Montalcini contract of MIUR. BJS was partially supported by the Oxford Martin School programme on Bio-Inspired Quantum Technologies and EPSRC grants EP/E036066/1 and EP/K034480/1.
Corresponding author: Hugo Defienne. Hugo.defienne@lkb.ens.fr

AUTHOR CONTRIBUTIONS

H.D., M.B. and S.G. conceived the experiment, with contributions from I.W. and B.S. H.D. performed the experiment and analyzed the results. All authors contributed to discussing the results and writing the manuscript.

ADDITIONAL INFORMATION

Correspondence and requests for materials should be addressed to H.D. and S.G.

COMPETING INTERESTS

The authors declare no competing interests.

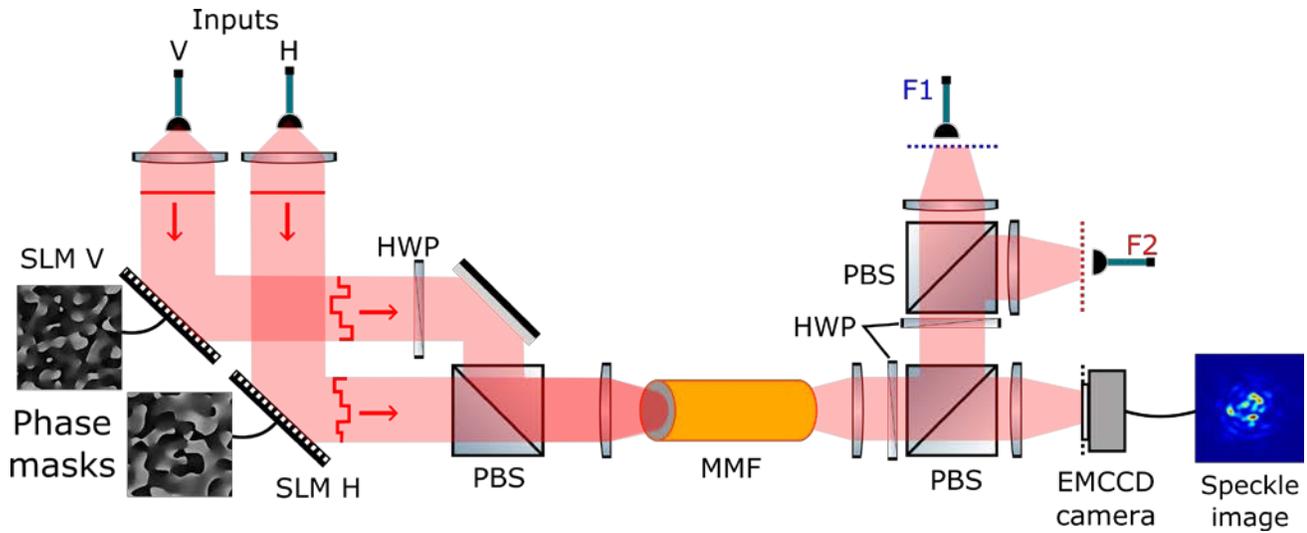

**Figure1: Apparatus to control the propagation of photon-pairs through a multimode fiber.**
Photon pairs from a SPDC source (not shown) are injected in a 50 μm diameter core graded index MMF with orthogonal polarizations. Two spatial light modulators (SLM H and SLM V) are used to shape the transverse spatial waveform of each photon. Output light is monitored by an EMCCD camera and two polarization maintaining single mode fibers (PMSMF) F1 and F2 connected to APDs all imaging the output plane of the MMF. A half wave plate (HWP) and polarizing beam splitter (PBS) positioned just after the MMF permits selection of one specific polarization of the output field.

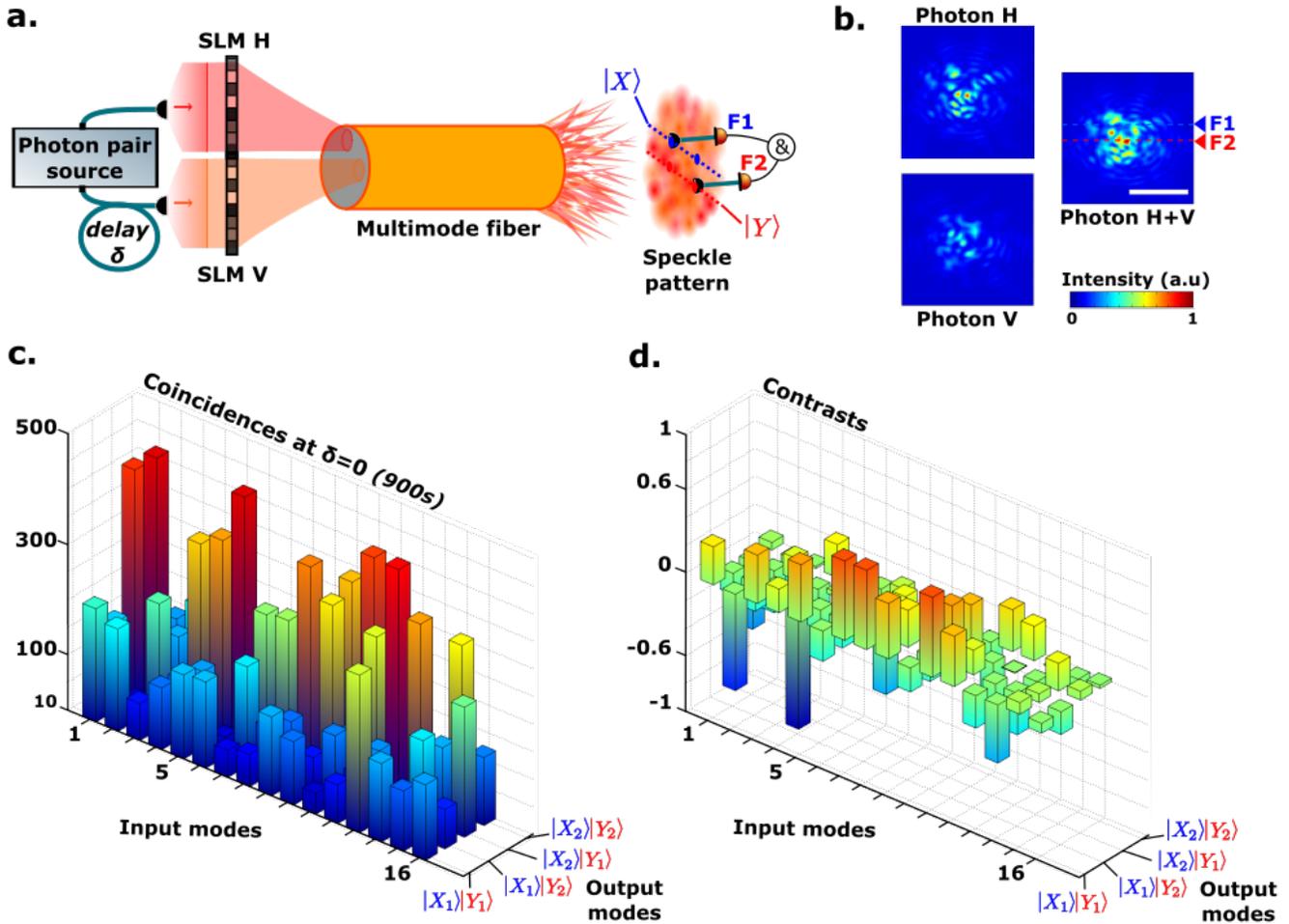

**Figure 2: Experimental measurement of the two-photon matrix.**
**(a.)** Schematic of the experimental setup using a two-photon state injected into two different transverse spatial modes. **(b)** Recorded intensity images of the corresponding output speckle patterns for photon H only, photon V only and H+V simultaneously, containing about 50 independent speckle grains. The scale bar represents 25μm in the output plane of the multimode fiber. The H+V speckle pattern corresponds to the incoherent sum of each individual case. **(c.)** Coincidence detection patterns between Fiber 1 (F1) and Fiber 2 (F2) reconstructed for 16 two-photon input states programmed by the SLMs. F1 and F2 scan four output coincidence positions denoted $|X_1Y_2\rangle, |X_1Y_2\rangle, |X_1Y_2\rangle, |X_1Y_2\rangle$. The $16\times 4$ coincidence matrix measured here represents a subset of the complete two-photon transmission matrix that comprises of approximately $(380\times 380)^2$ elements. **(d.)** For the same $16\times 4$ elements, differences observed in coincidence patterns measured with distinguishable (δ=0.4mm) or indistinguishable (δ=0) photons are quantified by reconstructing the non-classical contrast matrix. A contrast is given by the formula $C = (R_{\delta=0} - R_{\delta=0.4mm})/R_{\delta=0.4mm}$ where $R_\delta$ is the coincidence rate of a two photon output state at a path length difference of δ. This matrix is a clear signature of quantum interference effects occurring in the MMF. Coincidences are monitored for 900s with a coincidence window of 2.5 ns.

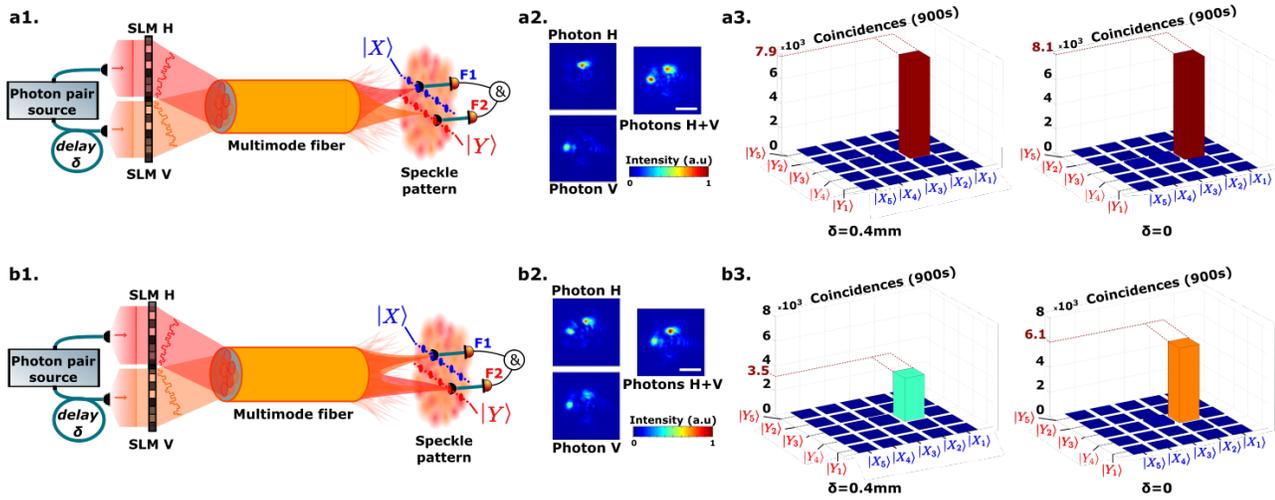

**Figure 3: Focusing photon-pairs in a targeted two-photon output state of the MMF**

The first configuration (**a1.**) directs photon H to output position $|X_2\rangle$ of F1 and photon V to output position $|Y_3\rangle$ of F2. This is visible in the direct images measured using the EMCCD camera (**a2.**). Coincidence profiles (**a3.**) are measured for 25 coincidence output fiber position pairs. Coincidence rates in the targeted two-photon state $\hat{a}^+_{|X_2\rangle}\hat{a}^+_{|Y_3\rangle}|0\rangle$ are about 100 times higher than the background, for both distinguishable and indistinguishable photons. The second configuration (**b1.**) corresponds to photon H being prepared in a superposition of output states $|X_2\rangle$ and $|Y_3\rangle$ with a relative phase $\varphi_H = 0$ and photon V directed to a superposition of the same output states with a relative phase $\varphi_V = 0$. Direct images measured with the EMCCD confirm that both photons are directed to the two output states (**b2.**). The effect of non-classical interference is shown on the output coincidence speckle patterns (**b3.**), where we observe an increase by 72% of the coincidences rate in the state $\hat{a}^+_{|X_2\rangle}\hat{a}^+_{|Y_3\rangle}|0\rangle$ in the indistinguishable case (δ = 0) compared to the distinguishable case (δ = 0.4mm). Coincidence measurements are acquired for 900s with a coincidence window of 2.5 ns. The white scale bars represent 25 µm in the output plane of the multimode fiber.

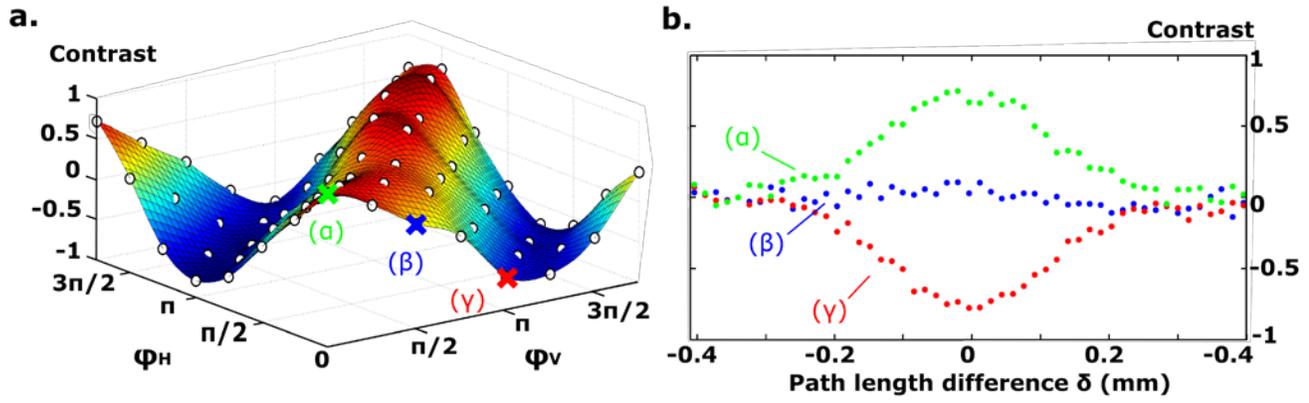

**Figure 4: Deterministic control of quantum interferences**
(**a.**) Non-classical interference contrast measured for two photons mapped onto a superposition of two output states with different phase settings $(\varphi_H, \varphi_V)$. The non-classical contrast is defined as $C = (R_{\delta=0} - R_{\delta=0.4mm})/R_{\delta=0.4mm}$ where $R_\delta$ is the two-photon coincidence rate of the targeted output state $\hat{a}^+_{|X_2\rangle}\hat{a}^+_{|Y_3\rangle}|0\rangle$ at a path length difference of δ. Contrast values are measured with 8 x 8 = 64 phase settings. (**b**) Contrast values for three phase settings (α): $(\varphi_H = 0, \varphi_V = 0)$, (β): $(\varphi_H = 0, \varphi_V = \pi/2)$ and (γ): $(\varphi_H = 0, \varphi_V = \pi)$, as a function of the path length difference between input photons δ. The observed effects are consistent with the initial indistinguishability of the photons evaluated with a HOM experiment (see Supplementary Materials). Data are acquired for 290s with a coincidence window of 2.5ns.